\documentclass{JHEP} 
\usepackage{amsmath}
\usepackage{amssymb,amsfonts}
\usepackage{epsfig}

\title{High Energy Scattering on Distant Branes}

\author{C. Bachas\\
 Laboratoire de Physique Th{\'e}orique
de l' Ecole Normale Sup{\'e}rieure\thanks{Unit{\'e} mixte  du
CNRS et de l ENS,  
UMR 8549.} \\ 
24  rue Lhomond, {}F-75231 Paris Cedex 05, France\\
\email{bachas@lpt.ens.fr}}

\author{B. Pioline\\
Centre de Physique Th{\'e}orique\thanks{Unit{\'e} mixte du
CNRS et de l' EP, UMR 7644}, \\ Ecole
Polytechnique, {}F-91128 Palaiseau, France\\
\email{pioline@cpht.polytechnique.fr}}

\abstract{We consider the elastic scattering of two open strings
living on two D-branes separated by a distance $r$. We compute the
high-energy behavior of the amplitude, to leading order in string
coupling, as a function of the scattering angle $\phi$ and of the
dimensionless parameter $v= r/(\pi\alpha^\prime\sqrt{s})$
 with $\sqrt{s}$ the
center-of-mass energy. The result  exhibits an interesting phase
diagram in the $(v,\phi)$ plane, with a transition at the production
threshold for stretched strings at $v=1$.  
We also discuss some more  general features of the open-string
semiclassical world-sheets, and 
 use T-duality to give a quantum tunneling interpretation
of the exponential suppression at high-energy.}

\preprint{CPTH-S726.0799\\LPTENS-99/26\\ \hepth{9909171}}

\begin{document}
\section{Introduction}

The purpose of the present work  is to discuss some aspects of the
high-energy fixed-angle scattering of open strings. 
In string perturbation theory such
scattering is exponentially soft  
and is dominated by smooth semiclassical 
trajectories  \cite{V,AAM,GM,GMa,B}.
From a technical viewpoint the external energy-momentum vectors
can be thought of as null Minkowski  charges  on the worldsheet, and 
the  saddle-point trajectories correspond to 
configurations of electrostatic equilibrium. Such configurations have
been exhibited at each order in the genus  expansion by Gross and
Mende \cite{GM}. They describe the elastic scattering of two  closed
untwisted strings  and also,  after appropriate identifications
\cite{GMa,B},  of two open strings or of a
single closed-string off a D-brane. The amplitude for all these fixed-angle
processes decays exponentially fast with the squared center-of-mass
energy ($s$) or momentum transfer ($t$). The suppression is, however,
less and less severe at higher orders, so that perturbation theory
will  eventually break down \cite{MO}. This is consistent with
the existence of some novel, possibly parton-like structure  at
substringy distances \cite{S} (see also \cite{DKPS,GG,GHV}).

    An interesting variation of the saddle-point problem arises when
the scattering strings are held a finite distance $r$ apart. 
This can be the distance between two D-branes in the case of open
strings, or between two orbifold fixed-points in the case of twisted
closed strings. For open strings, the
leading contribution to the scattering amplitude 
comes from worldsheets with cylindrical topology. The cylinder
stretches between the two distinct D-branes which, 
in the `electrostatic analog', behave like capacitor plates.  
Our main
technical  result in the present  work will be  the exact solution 
of this modified   electrostatic problem,  generalizing  the
result  of \cite{AAM,GMa}.

The relevant new parameter turns out to be $v=
r/(\pi\alpha^\prime\sqrt{s})$, in accordance with the naive expectation
that the size of fundamental strings grows with the center-of-mass energy
as $\sim \alpha^\prime\sqrt{s}$. We will show that the amplitude
has an interesting phase structure in the  $(v,\phi)$ plane, where
$\phi$ is the  angle of scattering. As $v$  ranges from infinity to zero 
 at fixed $\phi$, our  solution interpolates continuously  between massless closed-string
exchange and the 
exponentially-soft behavior   of \cite{AAM,GMa}. The electrostatic
equilibrium becomes unstable at $v<1$,  reflecting the fact that we have
crossed the threshold for production of
two massive stretched strings at an intemediate state. 
This qualitative change of behavior may  be of phenomenological relevance, particularly
in the context of 
low-string-scale models, in which the threshold $v=1$ for nearby
branes could be in the few-TeV region.

The exponential suppression of  the perturbative high-energy amplitudes
is reminiscent of a quantum tunneling process, and also  of
a minimal  surface (soap-bubble)  problem.  T-duality in open-string
theory makes these interpretations precise. Open string theory allows
furthermore a better understanding of the relation between
`electrostatic stability' on the world sheet and unitarity cuts of the
corresponding amplitude. These simple comments 
 have not been discussed to our
knowledge previously in the literature, and  shed
interesting new light on 
the physics of high-energy collisions.  We have  devoted a
(non-technical)  part of our paper to explaining them.

   The plan of the paper is as follows: in section 2 we review the
saddle-point calculation of the disk amplitude, with an emphasis on
topological features  of the semiclassical trajectory that generalize
to higher orders. In sections  3 and 4 we set up and solve the saddle point
problem on the annulus, and discuss the phase diagram of the amplitude
in the $(v,\phi)$ plane. We end with some comments on T-dual
interpretations of the high-energy collision process.


\setcounter{equation}{0}
\section{ Topology of Semiclassical Trajectories}

We consider the scattering of two incoming open-string states (1 and 2)
into two outgoing open-string states (3 and 4), with  Mandelstam
variables  $s = -(p_1+p_2)^2$, $t = -(p_1+p_3)^2$ and   $u = -(p_1+p_4)^2$.
The high-energy fixed-angle behavior of the amplitude 
does not depend on the
details of the external states, provided  all  masses are  kept finite
in the limit.  We can assume thus for
simplicity that all the external states are massless. 
In terms of the  center-of-mass energy $E$ and scattering angle $\phi$
the Mandelstam variables read 
\begin{equation}
s= 4E^2\ , \ \ -{t\over s} = \sin^2{\phi\over 2}\ ,
\ \ {\rm and} \ \  -{u\over s} = \cos^2{\phi\over 2}\ ,
\end{equation}
where $\phi$ ranges from $0$  for forward
scattering to $\pi$ for backward scattering. Note that in the physical kinematic
region $s$ is positive while $t$ and $u$ are negative.
Momentum conservation implies that $s+t+u = 0$.

The high-energy behavior at  tree-level  can be 
extracted  from  the  Veneziano  formula, 
\begin{equation}
{\cal A}_{\rm disk}^{(oo\rightarrow oo)}\sim \frac{\Gamma(-\alpha't)\Gamma(-\alpha'u)}
{\Gamma(\alpha's)} + (s\leftrightarrow t) + (s\leftrightarrow u)\ .
\label{eq:ven}
\end{equation}
The three terms in the above amplitude correspond to 
different topological arrangements  of the external states, and are multiplied by the
Chan-Paton factors  
 ${\rm tr}(1324)$, 
${\rm tr}(1234)$ and 
${\rm tr}(1243)$ respectively.
Using Stirling's approximation for the Gamma functions,
  $\Gamma(x)\sim \sqrt{2\pi} x^{x-1/2} e^{-x}$,
one finds
\begin{equation}  
\ {\cal A}_{\rm disk}^{(oo\rightarrow oo)}
 \sim e^{- \alpha^\prime{ \tilde{\cal E}}_{\rm disk}   }\  
\end{equation}
where the exponential weight is given by 
\begin{eqnarray}
\label{eq:ven1}
\tilde{\cal E}_{\rm disk}  &=& 
 s \log\vert s\vert +t \log\vert t\vert
+  u \log\vert u\vert \\
&=& -{s} \left[  \cos^2{\phi\over 2}
\log\left(\cos^2{\phi\over 2}\right)+ \sin^2{\phi\over 2}
\log\left(\sin^2{\phi\over 2}\right)  \right]
\nonumber
\end{eqnarray}

Strictly-speaking only the first of the three terms in \eqref{eq:ven}
has a well-behaved  asymptotic limit in the physical kinematic
region. The other two  have a
series of poles at integer values of $\alpha's$,  corresponding to the
excited  open strings that can propagate on-shell   in the intermediate
channel. For these singular amplitudes, the asymptotic limit
refers to the
residue at the poles,  or to the  imaginary (absorptive) part averaged over some
range of incoming momenta. Note that
the  amplitude without s-channel poles is obtained
 when  the incoming strings
interact by joining 
\FIGURE{\begin{picture}(300,150)(0,0)
\put(50,0){\epsfig{file=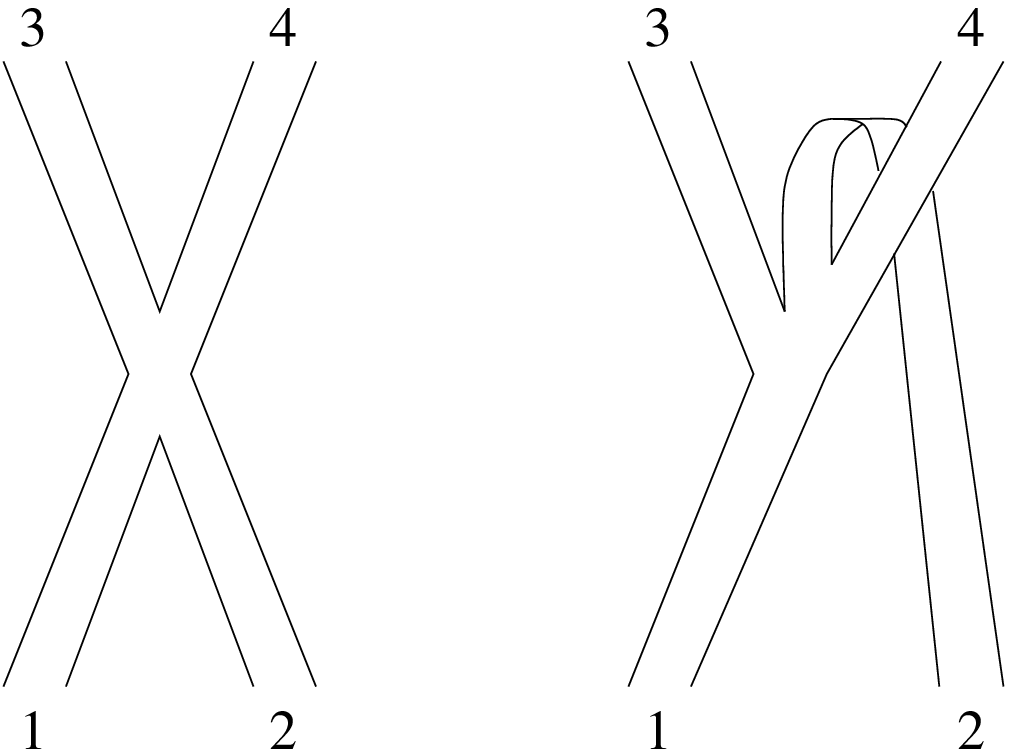,height=5cm}}
\end{picture}
\caption{Open-string 
 amplitudes with (left) or without (right) s-channel poles\label{vert}}}
and splitting in the middle, as 
illustrated in figure \ref{vert}. Note also that the asymptotic limit has  forward/backward
symmetry, as can be seen from the fact that  $\phi\leftrightarrow \pi-\phi$. 
amounts to exchanging  the Mandelstam variables $t$ and $u$.

 It is instructive to  rederive the above  results
 by a saddle point calculation, in which one
extremizes the `electrostatic energy' of four null Minkowski  charges placed at the
boundary of a disk. The advantage of the `electrostatic' calculation
 is that it can be extended
to  more general situations  when the amplitude --  an integral over positions of
the vertex operators and moduli of the world-sheet -- does not admit  a
closed-form expression.
Let us represent  the  disk as the upper-half plane, and let $x_i$ be
the points of insertion of the vertex operators on the boundary. 
The Veneziano amplitude is given by an  integral 
 \begin{equation}
\ {\cal A}_{\rm
disk}^{(oo\rightarrow oo)}
\sim \int d\zeta\  e^{- \alpha^\prime{{\cal E}}_{\rm disk}}\ , 
\end{equation} 
 where $\zeta = (x_1-x_3)(x_2-x_4)/(x_1-x_2)(x_3-x_4)$ is  the $Sl(2,R)$
invariant cross-ratio, and 
the `electrostatic energy'  reads
 \begin{equation}
{\cal E}_{\rm disk} = 
 t \log \vert\zeta\vert  + u \log\vert
  1-\zeta\vert\ .
\label{eq:ven2}
\end{equation} 
If we use the $SL(2,R)$ symmetry to fix $x_1=0$, $x_2=1$ and
$x_4=\infty$, then $\zeta = x_3$.
The (1324), (1234) and (1243) permutations of external states
 correspond therefore to the integration intervals 
$(0,1), (1,\infty)$ and  $(-\infty,0)$. The first of these three
integrals is as expected real and convergent, while the other two 
diverge as  $\vert\zeta\vert\to\infty$, 
and must be defined by analytic continuation.

The electrostatic energy ${\cal E}_{\rm disk}$ 
has a  saddle point  at 
\begin{equation}
{\tilde \zeta} = -{t\over s} = \sin^2{\phi\over 2}\ ,
\label{eq:ven3}
\end{equation}
where it  takes precisely the value in \eqref{eq:ven1}. 
Since  this is the only 
saddle point in the complex $\zeta$-plane,  we expect it to govern  the 
behavior of the amplitude for any ordering of the external states.
What  is important,  however,  to
realize is that the topology of the semiclassical trajectory is
always of the  (1324) type --
 ${\tilde \zeta}$ goes indeed from
0 to 1 as the scattering angle ranges from 0 to $\pi$. 
 This is not of
course surprising~: a smooth  semiclassical trajectory of the 
(1234) or  (1243) type  would have been  incompatible with the fact that
highly-excited open strings can propagate  on-shell at an
intermediate stage. In the language of the  electrostatic analog, the two
incoming particles `attract' so that  they cannot  
equilibriate  if they are  adjacent. Note furthermore that 
the configuration
\eqref{eq:ven1} is one of   stable equilibrium, as should be
expected of a saddle point which  governs the behavior of an integral
that is both real and convergent. 
\footnote{The equilibrium 
is on the other hand  unstable if we allow the 
charges to  move off the real axis. This
reflects the fact that in closed-string amplitudes one  cannot
suppress the s-channel poles by a suitable choice of Chan-Paton factors.}

\EPSFIGURE{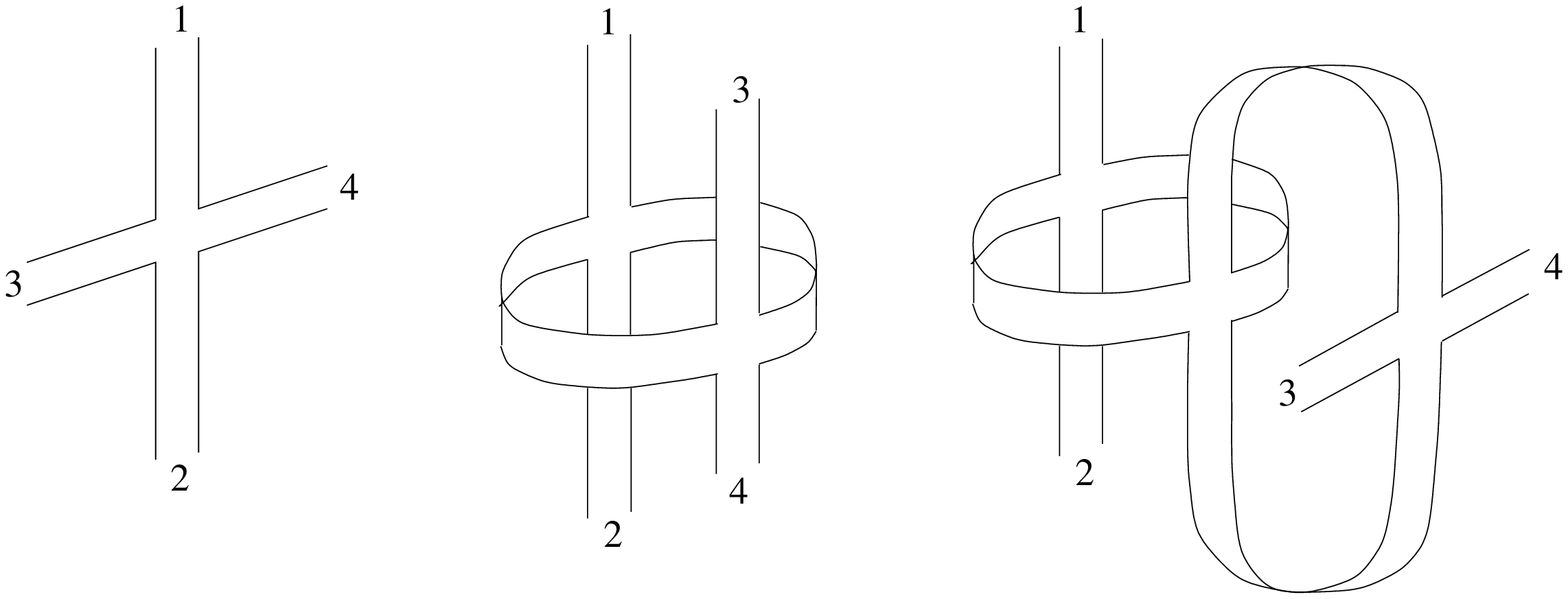,height=4cm}{Topology of higher-genus 
saddle points \label{nonp}} 

     Extending these arguments to higher orders, leads us to expect 
that the saddle-point trajectories have a topology forbidding
one-particle intermediate states to go  on-shell. This can be achieved
by insisting that successive scatterings are of the non-planar type 
shown in figure \ref{vert}.
Two such successive
scatterings  correspond, in particular,  to the non-planar annulus diagram, three
scatterings to the disk diagram with a closed-string  
handle etc (see figure \ref{nonp}). More generally for any odd (or even) number   N of
scatterings one finds a
world-sheet topology
with one (or two) boundaries and  $[{N-1\over 2}]$  closed-string
handles.  That the relevant semi-classical trajectories should be of this type 
agrees with the observation of Gross and Manes \cite{GMa} that, in the
corresponding electrostatic problem, empty world-sheet boundaries are
`shrunk away to zero by electrostatic pressure'.

\setcounter{equation}{0}
\section{Deformed Gross-Mende Saddle Points}
The elastic scattering discussed in the previous section assumed
massless open-strings living on a set of coincident D-branes. 
We want now to analyze what happens when the scattering strings live
on two D-branes separated by a distance $r$. The kinematics of one
such process is illustrated in figure \ref{nonp}. The two
incoming open strings (1 and 2) live  on two distant  D-branes, as do the
two outgoing strings (3 and 4). The amplitude, proportional to the
Chan-Paton factor ${\rm tr}(13){\rm tr}(24)$, receives contributions from 
world-sheets with  at least two  boundaries, so that  the leading perturbative
contribution comes from the cylinder (or annulus) diagram. 
The  topology of figure \ref{nonp} is  special in that it does  not allow
one-particle unitarity cuts. 
The (12)(34) topology, by  contrast, has both  open- and closed-string 
channel poles,  the topologies with three insertions on a  boundary
have intermediate open-string poles, while the topologies with an
empty boundary allow a massless closed-string to propagate at zero
momentum down the throat. Following our discussion in the previous
section,  we should therefore expect the relevant semiclassical trajectories
to have  the  (13)(24) or the  (14)(23) topologies. 

\FIGURE{
\begin{picture}(300,150)(0,0)
\put(0,20){\epsfig{file=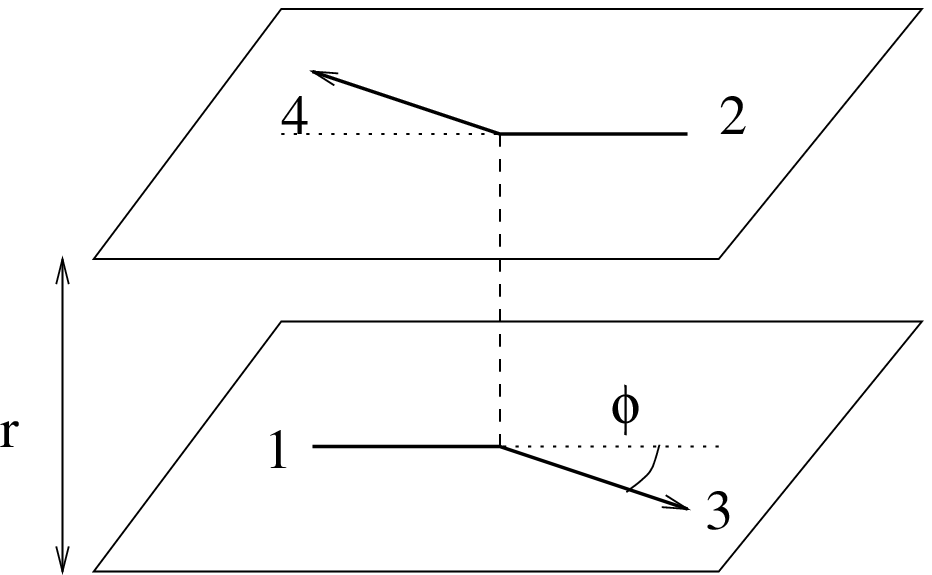, height=3cm}}
\put(180,0){\epsfig{file=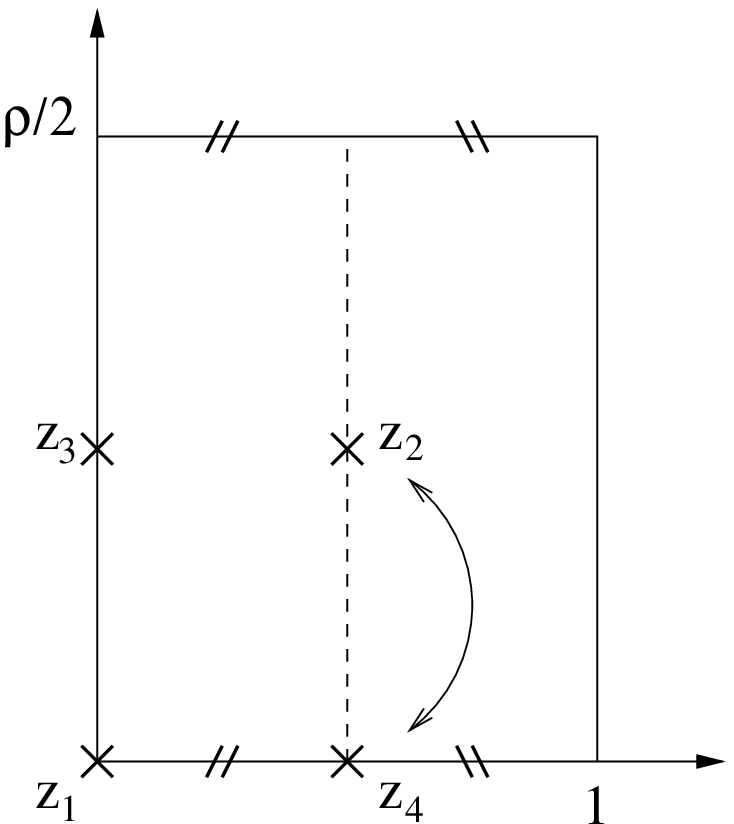,height=5cm}}
\end{picture}\caption{Kinematics and 
saddle point configurations  on the annulus\label{ann}}}

Let us set up now the corresponding electrostatic problem. 
We consider the annulus as a (doubling)
torus of modulus $\tau = i\rho/2$, 
modded out by a reflection symmetry along the imaginary axis. 
The bosonic propagator for fields on the boundaries  reads
 \begin{equation}
G(z\vert\tau) = \log\chi \ , \ \  
\chi = 2\pi e^{-\pi y^2/\tau_2}\Bigl\vert
{\theta_1(z\vert\tau)\over \theta_1^\prime(0\vert\tau)}
\Bigr\vert\ \ , 
\label{eq:prop}
\end{equation}
where $z=x+iy$ and $\theta_1^\prime=2\pi \eta ^3$.
Using the momentum conservation and mass-shell conditions 
one finds the following expression for the  `electrostatic energy' 
\begin{align}
\label{eq:energy}
{\cal E}_{\rm annulus} =  {r^2\rho \over 4\pi{\alpha^\prime}^2} &+  s \left[
 \log\vert
\theta_1(z_{12})\theta_1(z_{34})\vert - 2\pi (y_{12}^2+y_{34}^2)/\rho 
\right]
\nonumber\\
& +t \left[\log\vert
\theta_1(z_{13})\theta_1(z_{24})\vert - 2\pi
 (y_{13}^2+y_{24}^2)/\rho\right]
\\
& +u\left[ \log\vert
\theta_1(z_{14})\theta_1(z_{23})\vert - 2 \pi
(y_{14}^2+y_{23}^2)/\rho\right]
\nonumber
\end{align}
Here the $\rho$-dependence
of the Jacobi functions has been suppressed, and $z_{ij} = z_i-z_j$ is
the relative position of the inserted vertex operators, which obeys 
${\rm Re}z_{ij}= 0, 1/2$  for $i,j$ lying on the same or opposite boundaries.
The amplitude is given by an integral  over  the positions  $z_i$ of the `charges'  and over
the  modulus $\rho$ of the world-sheet, 
\begin{equation}
\ {\cal A}_{\rm
annulus}^{(oo\rightarrow oo)}
\sim \int d\rho\ \prod_i \oint dz_i\;
  e^{- \alpha^\prime{{\cal E}}_{\rm annulus} }\ . 
\label{eq:intt}
\end{equation}

   The novel feature in expression \eqref{eq:energy} is the 
 contribution to the `electrostatic energy' due to  the stretching of the
world-sheet between the two branes. It  can be thought of as the
 energy of a constant
field created by a fixed  potential drop between the two boundaries
of the surface. A simple scaling argument shows that the relevant
new parameter is 
\begin{equation}
v = { r \over \pi{\alpha^\prime}\sqrt{s} }\  .
\label{eq:param}
\end{equation}
This can be interpreted  as the (inverse) ratio of  the
effective size of the incoming  strings, which grows  as $\sim
\alpha^\prime\sqrt{s}$ at high energies, to the distance $r$ between the
branes.  Alternatively, it is
 the mass of a pair of 
strings that stretch  between the two D-branes, 
divided by  the total center-of-mass energy. The
significance of these facts
 will become apparent in the following section.

    The saddle points of ${\cal E}_{\rm annulus}$ are harder to
analyze than those of ${\cal E}_{\rm disk}$. Following
refs. \cite{GM,GMa} we  focus our attention to  configurations in which
the  four charges are placed  half-a-cycle apart on the doubling torus. 
Symmetry ensures that  the energy is then automatically extremized  under
variations of the $z_i$. 
There are two such configurations with the topology (13)(24), 
\begin{align}
\label{eq:insertions}
& ({\rm A}): \ \ {\tilde z}_1=0\ , \quad {\tilde z}_2 = {1\over
2}+{i\rho\over 4} \ ,\quad 
{\tilde z}_3 = {i\rho\over 4}\ ,\quad {\tilde z}_4= {1\over
2}\ , 
\nonumber\\
{\rm and}\ \ \ \ \ \ \ \ \ \ &
\\
& ({\rm B}): \ \ {\tilde z}_1=0\ , \quad {\tilde z}_2 = {1\over 2}\ ,\quad 
{\tilde z}_3 = {i\rho\over 4}\ ,\quad {\tilde z}_4= {1\over 2}+{i\rho\over 4}\ ,
\nonumber
\end{align}
(see figure \ref{ann}). 
 Using standard  properties of the Jacobi $\theta$ functions  we
find the following electrostatic energies  for them:
 \begin{equation}
{ {\cal E}}^{(A)} = 
{r^2\rho\over 4\pi{\alpha^\prime}^2}  + 2 s  \log \vert \theta_3 \vert  + 
2 t  \log\vert\theta_4\vert + 2 u  \log\vert\theta_2\vert \ ,
\label{eq:energ}
\end{equation}
and 
 \begin{equation}
{ {\cal E}}^{(B)} = 
{r^2\rho\over 4\pi{\alpha^\prime}^2}  + 2 s  \log \vert\theta_2\vert + 
2 t  \log\vert\theta_4\vert + 2 u  \log\vert\theta_3\vert \ , 
\label{eq:energ1}
\end{equation}
where the $\theta$-functions are  evaluated
at zero argument. The problem is now  reduced to that of finding 
 extrema of these energies with respect to  the surface modulus
 $\tau= i\rho/2 $.

  We will here  solve this mathematical problem
analytically -- the physical implications of our results will be
discussed in the following section. It will be enough  to analyze
 configuration (A),  since for
 (B) we  need only  interchange the roles of $s$ and $u$
 in our final expressions. 
Extremizing the energy \eqref{eq:energ}
with respect to $\rho$ gives the equation
\begin{equation}
{r^2\over 8\pi{\alpha^\prime}^2}+  
s {\theta_3^\prime\over \theta_3}+ t {\theta_4^\prime\over \theta_4}+
u {\theta_2^\prime\over \theta_2}  = 0
\label{eq:saddle1}
\end{equation}
where  $\theta_j^\prime \equiv \partial_\rho\theta_j ={i\over 2}
\partial_\tau\theta_j$. Let us first recall how this equation is
solved  in  the case of coincident branes  \cite{GMa}. 
The trick is to identify the Mandelstam variables with the three terms
of the  Riemann identity
$\theta_3^4-\theta_4^4 - \theta_2^4 = 0$. It is indeed straightforward
to check that the ansatz 
\begin{equation}
 s = a {\theta_3}^4 \ , \ \  t= -a {\theta_4}^4\ , \ \ 
u = -a {\theta_2}^4 \ .
\label{eq:sol1}
\end{equation}
solves both the constraint $s+t+u=0$, and the 
saddle-point equation when   $r=0$. 
The saddle-point modulus ${\tilde\tau} = i{\tilde\rho}/2$ is, in this
special case, determined implicitly by the Picard map,   
\begin{equation}
-{u\over s} = \cos^2{\phi\over 2} =
 \left({\theta_2({\tilde\tau})\over
 \theta_3({\tilde\tau})}\right)^4\ , 
\label{eq:exp2}
\end{equation}
which is one-to-one from the fundamental domain for the modular group
$\Gamma(2)$ to the complex plane. The modulus ${\tilde\tau}$ is hence
a function of the scattering angle only, as expected. 
The energy at the saddle point takes the value
\begin{align}
2 { \tilde{\cal E}^{(A)}\Bigl\vert_{v=0}} & = 
 s \log\vert s\vert +t \log\vert t\vert
+  u \log\vert u\vert \\
& = -{s} \left[  \cos^2{\phi\over 2}
\log(\cos^2{\phi\over 2})+ \sin^2{\phi\over 2}
\log(\sin^2{\phi\over 2})  \right] \ ,
\label{eq:exp1}
\end{align}
which is exactly half the electrostatic energy on the disk, in
agreement with the one-loop result of Gross and Manes \cite{GMa}. 
Intuitively the exponential suppression is less severe because the
total momentum transfer is split between two equally-softer scatterings.
Note incidentally that, at
fixed collision energy, the maximal suppression is attained at right-angle  scattering 
and corresponds to a 
square semiclassical world-sheet ($\tilde\tau=i$).

Extending  the above trick
to  non-zero $r$  requires a modular  identity
whose  three terms add up to a non-vanishing  constant. Such an
identity, well 
known from the study  of BPS saturated  string amplitudes, is
\begin{equation}
 \Bigl[ \theta_3^2 \theta_3' \theta_3'
 -\theta_4^2 \theta_4'\theta_4'-  
\theta_2^2 \theta_2'\theta_2'\Bigr]/\eta^{12}  = - {\pi^2 \over 4}\ .
\label{eq:id}
\end{equation}
Using this and  the second derivative  of the Riemann identity it can
 be checked  that 
\begin{equation}
 s = a {\theta_3}^4 +  b
{\theta_3^\prime\theta_3^3 \over \eta^{12}} 
\ , \ \ -t= a {\theta_4}^4 + b
{\theta_4^\prime\theta_4^3 \over \eta^{12}}\ , \ \
 -u = a {\theta_2}^4+ b 
{\theta_2^\prime\theta_2^3 \over \eta^{12}} \ ,
\label{eq:sol2}
\end{equation} 
solves the  constraint and 
the saddle-point  problem \eqref{eq:saddle1} provided one chooses 
\begin{equation}
 b = {r^2\over 2\pi^3 {\alpha^\prime}^2}\equiv \frac{s v^2}{2\pi}\ .
\label{eq:sol22}
\end{equation}

It will be useful  to recast the above solution in a form that
expresses the  saddle-point modulus $\tilde \rho$
as   a  function of $v$ and of the scattering angle
$\phi$. Eliminating $a$ gives  the implicit relation
\begin{equation}
-{u\over s} = \cos^2{\phi\over 2} =
 \left({\theta_2\over
 \theta_3}\right)^4\times 
\left[ 1 + {\theta_3^4 v^2 \over 2 \pi  \eta ^{12}  }
\left( {\theta_2^\prime \over \theta_2}-{\theta_3^\prime \over
 \theta_3}
\right) \right]\ .
\label{eq:exp3}
\end{equation} 
This can be further simplified with the help of the  identities~:
\begin{subequations}
\begin{eqnarray}
{\theta_2^\prime \over \theta_2} &=& -{\pi\over 24} \left(
E_2+\theta_3^4 + \theta_4^4 \right) \\
{\theta_3^\prime \over \theta_3} &=& -{\pi\over 24} \left(
E_2+\theta_2^4 - \theta_4^4 \right) \\
{\theta_4^\prime \over \theta_4} &=& -{\pi\over 24} \left(
E_2-\theta_2^4 - \theta_3^4 \right) 
\end{eqnarray}
\end{subequations}
where $E_2=-(24/\pi)\partial_\rho \log\eta$ 
is the holomorphic weight-2 Eisenstein series. Some
straightforward algebra leads to the following
relation between $\bar\rho$, $v$ and $\phi$,
\begin{equation}
({\rm A})~: \ \ \ -{u\over s} = \cos^2{\phi\over 2} =
{\theta_2^4(\bar\tau) - v^2\over
\theta_3^4(\bar\tau)}\ \ .
\label{eq:exp5}
\end{equation}
This generalizes the Picard map \eqref{eq:exp2} to the case of
non-coincident branes. For later convenience we write explicitly also
the relation obtained for the configuration in which positions $z_2$
and $z_4$ have been exchanged. Exchanging $s$ and $u$ (and remembering
that $s$ entered also in the definition of $v$) we find
\begin{equation}
({\rm B})~: \ \ \ -{u\over s} = \cos^2{\phi\over 2} =
{\theta_3^4(\bar\tau) - v^2\over
\theta_2^4(\bar\tau)}\ \ .
\label{eq:exp55}
\end{equation}
As will be verified numerically in the coming section, these equations
give {\it all} the extrema of ${ {\cal E}}^{(A)}$ and ${ {\cal E}}^{(B)}$
on the  real positive $\rho$-axis. We will also see that depending on
the values of $\phi$ and $v$, eq.  \eqref{eq:exp55} has one, two or zero
real solutions -- in the latter case this  means that our formal ansatz was
inconsistent.

 We  close this  section with some remarks on other possible
saddle points. First, in addition to  A and B, there are  four
more configurations of electrostatic equilibrium, at fixed $\rho$,
corresponding to  inequivalent permutations of the charges. The extrema
$\tilde \rho$ of
their  energy must satisfy  an equation obtained from \eqref{eq:exp5} 
after an appropriate permutation of the Mandelstam variables. The
(14)(23) extrema are in fact the same as  A and B,  up to an exchange of $\phi$
with $\pi -\phi$. For the (12)(34) configurations,  on the other hand,
 the $\rho$-extremization  has no solution, because  the  attracting
pairs of charges tend to shrink the boundaries of the annulus to a point. The
corresponding amplitude must be obtained by deforming the $z_i$
integration contours, as was the case for the
resonnant  amplitudes at  disk-level.  A harder question concerns
saddle points in the complex $\tau$-plane, which   become as we will
see relevant
in kinematic regions  where the (13)(24) amplitude must be
analytically continued.  For $v=0$, all saddle points are modular
transform of one another, and yield the same energy as $A$.
For $v>0$ however, modular invariance is broken and the degeneracy
between these saddle points is lifted. 

\setcounter{equation}{0}
\section{Phase Diagram of the Amplitude}

   We now want to use the above results in order to evaluate the
amplitude for the scattering process (13)(24) in the 
asymptotic high-energy limit. To this end let us first consider  the
potential singularities of the integral  \eqref{eq:intt}. Its
three possible degeneration limits,
\begin{equation}
 (a)\  \ z_{13} \ {\rm or} \ z_{24}\to 0\ , 
\ \ (b)\  \ \rho\to 0\ , \ \ {\rm and}  \ \ (c)\ \  \rho\to \infty \  ,
\label{eq:limits}
\end{equation}
correspond  to an intermediate
 (a) open string, (b) closed string,  or (c) pair of
stretched strings going on shell. We have already argued that the
first two singularities are absent because  the momentum flowing out of a
boundary is space-like. It can be indeed verified immediately that the
 electrostatic energy \eqref{eq:energy}  goes  to $+\infty$ in the
 limit (a), as long as the momentum transfer $t$ stays negative.  
To analyze the other limits one should note that, at fixed $\rho$,
configuration B is a global minimum of the energy. This is because
 the attracting (repelling) charges are as close (as far) 
 from each other as they could be. We need therefore only check whether the
 energy of this configuration stays
 bounded from below as $\rho$ varies.  Using the asymptotic
 properties of $\theta$ functions, and the definition
  \eqref{eq:param} of the parameter $v$,  we find~:
 \begin{equation}
{ {\cal E}}^{(B)}\ 
 \stackrel{\rho\to 0}{\sim}\  -\frac{\pi t}{\rho}+ 2t\log 2 + \frac{\pi v s}{4}\rho +\cdots
\label{eq:asym3}
\end{equation}
 \begin{equation}
 { {\cal E}}^{(B)}\ 
\stackrel{\rho\to \infty}{\sim}\  \frac{\pi s}{4}
\left(v-1\right)\rho +2s\log 2+\cdots \ ,
\label{eq:asym2}
\end{equation}
where the dots denote exponentially-small corrections. 
The limit (b) is thus energetically forbidden,  as  expected, 
while in the limit (c)
the  energy stays  bounded from below if and only if $v>1$. 
The reason for this  is easy to understand~:  $v=1$ is the
critical separation  at which the available center-of-mass energy suffices to
create a pair of open strings stretching between the two branes. The
instability of the electrostatic problem for $v<1$ signals therefore 
the opening  of a two-particle production threshold, 
and the appearance of an
imaginary part to the amplitude.

\FIGURE{
\begin{picture}(250,440)(0,0)
\put(5,320){\makebox{\epsfig{file=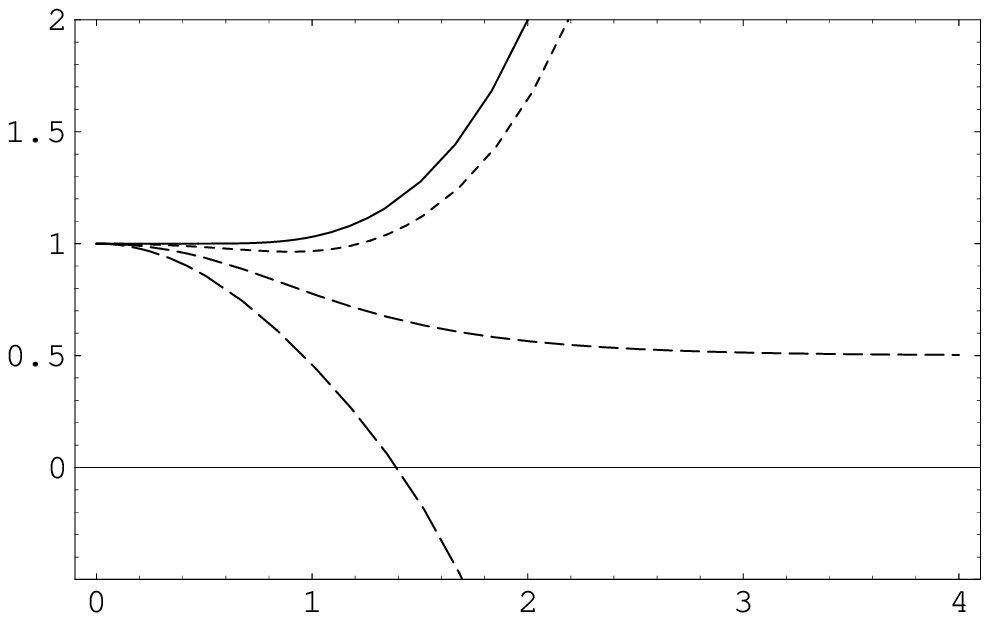,height=5cm}}}
\put(5,160){\makebox{\epsfig{file=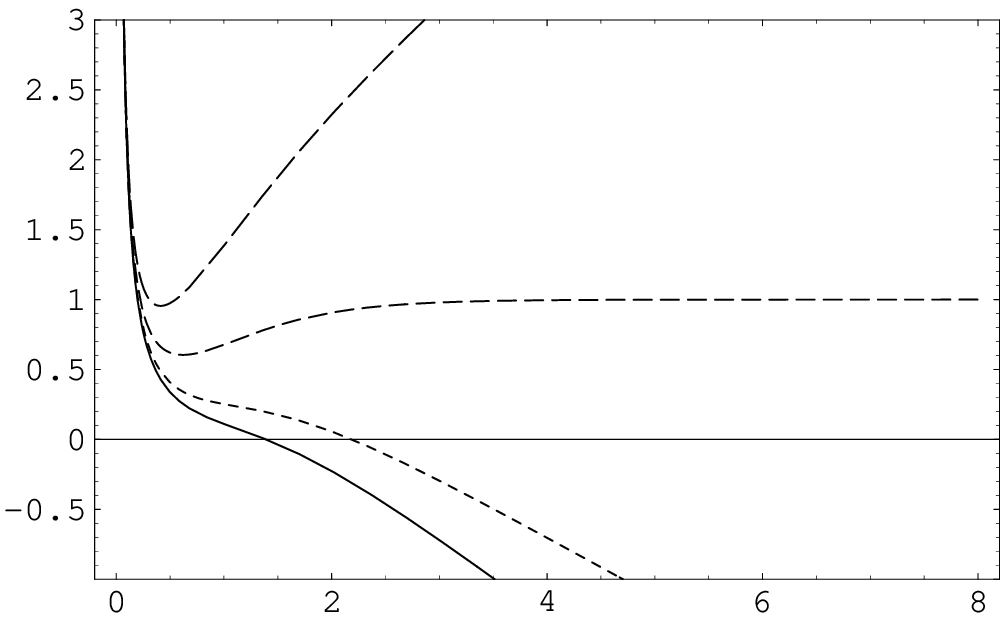,height=4.9cm}}}
\put(5,0){\makebox{\epsfig{file=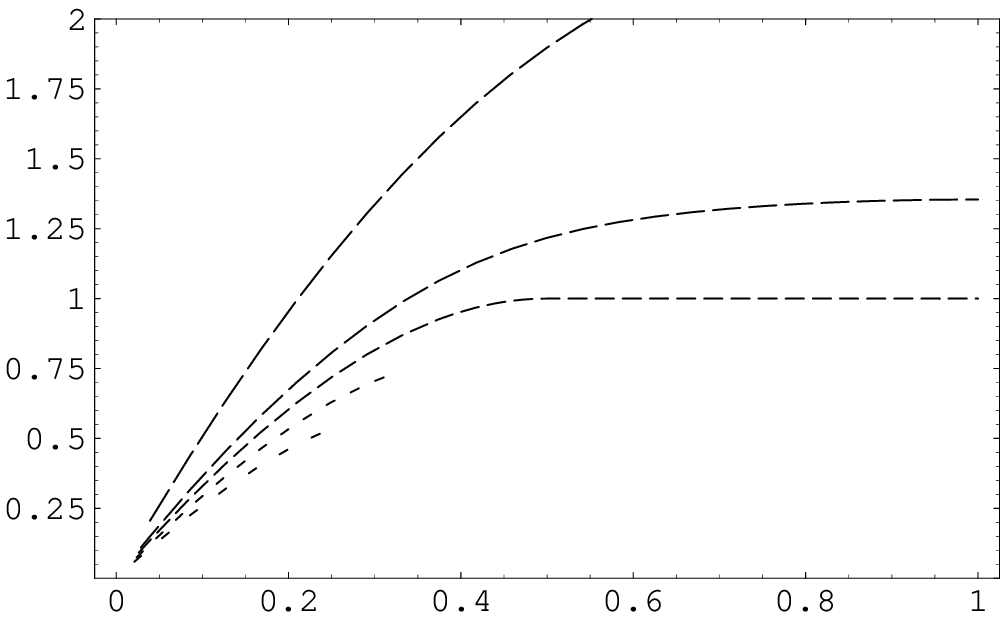,height=4.9cm}}}
\put(-30,430){\makebox{$\cos^2{\phi\over 2}$}}
\put(-50,276){\makebox{${\cal E}/{2s\log 2}$}}
\put(-50,120){\makebox{${\tilde{\cal E}}/{2s\log 2}$}}
\put(242,325){\makebox{$\rho$}}
\put(242,165){\makebox{$\rho$}}
\put(242,5){\makebox{$\phi/\pi$}}
\put(85,440){\makebox{$v=0$}}
\put(135,440){\makebox{$v=.5$}}
\put(200,388){\makebox{$v=1$}}
\put(60,340){\makebox{$v=1.5$}}
\put(70,122){\makebox{$v=1.5$}}
\put(185,102){\makebox{$v=1.1$}}
\put(185,77){\makebox{$v=1.0$}}
\put(110,55){\makebox{$v=0.9$}}
\put(90,40){\makebox{$v=0.8$}}
\end{picture}
\caption{Deformed saddle point B. Top: $\cos^2{\phi\over 2}$ vs. modular
parameter $\rho$ as in \eqref{eq:exp55}. Middle: electrostatic energy
vs. modular parameter $\rho$ at small scattering angle $\phi=\pi/10$.
Bottom: electrostatic energy at the minimum vs. angle. The solid
line corresponds to coincident branes, and the dashed  lines
to separated branes \label{figNS}.}
}

Since,  for $v>1$, the integral \eqref{eq:intt} is  real and 
convergent, it must be  dominated by the  minimum  `electrostatic energy'
in the integration range, i.e. by  ${\cal E}^{(B)}(\tilde \rho)$ at 
$\tilde\rho$  given 
implicitly by \eqref{eq:exp55}. We have solved this equation
numerically, and have plotted the results in figure \ref{figNS}. Note first that
the solution to the equation is unique, and that it corresponds indeed to a
global minimum. The two interesting limits to consider are (i)
the limit of large brane separation ($v\to\infty$) and (ii) the
approach to the stretched-string production threshold ($v\to 1_{+}$). 
In the $v\to\infty$ limit, the
saddle point approaches zero  as ${\tilde\rho}\sim
2\pi\alpha^\prime\sqrt{-t}/r$. and the amplitude takes the asymptotic form
\begin{equation}  
\ {\cal A}^{(oo\to oo)}_{\rm annulus}\  \stackrel{v\to \infty}{\sim}\
\  e^{- r\sqrt{-t} -  2t\alpha^\prime\log 2   }\ 
.  
\end{equation}
The leading exponential behavior is characteristic of the 
exchange of massless closed-string modes between the branes,  
\begin{equation}  
e^{- r\sqrt{t}} \sim \int d{\vec p}\; {e^{-i{\vec p}\;{\vec r}}\over {\vec
p}^{\; 2}+t}\ ,
\end{equation}
with  ${\vec p}$  the transverse momentum of the closed string,  
which is not conserved at
each individual vertex. The subleading term in the exponential is more
intriguing. A simple calculation shows that it cannnot be accounted
for by the
exponentially-rising density of intermediate states -- it represents  thus an
effective exponential enhancement (since $t$ is positive) of the off-shell vertices. 
Consider next the $v\to 1_{+}$ limit.
 The imaginary (absorptive) part of the amplitude is easy to calculate
at threshold, because the two stretched strings are in their ground
state and at rest. 
We will see  in the following section that the production of  two
large unexcited stretched  strings at rest in a  collision
is a process T-dual to  elastic right-angle scattering. This implies that
\begin{equation}  
\ {\rm Im}  {\cal A}^{(oo\to oo)}_{\rm annulus}\ \  \stackrel{v\to
1_{-}}{\sim}\ \ 
\ \Bigl\vert {\cal A}^{(oo\to oo)}_{\rm disk}(\pi/2) \Bigr\vert^2 \sim 
 e^{-  2s \alpha^\prime\log 2}\ 
. 
\label{eq:abs} 
\end{equation}
Let us compare this to the electrostatic energy plotted for $v=1_{+}$
in figure \ref{figNS}. As $\phi$ goes from 0 to $\pi/2$  the
minimum of the energy rises from 0 to $2s \log 2$, 
where it stays frozen while  the angle of scattering continues to vary between
$\pi/2$ and $\pi$. Thus the absorptive part saturates the amplitude
for backward scattering, while it is subleading when compared to the
real part for forward scattering. The semiclassical trajectory for
backward scattering becomes singular (${\tilde\rho}\to\infty$) at
the threshold as expected.  

\FIGURE{
\begin{picture}(300,180)(17,0)
\put(50,5){\epsfig{file=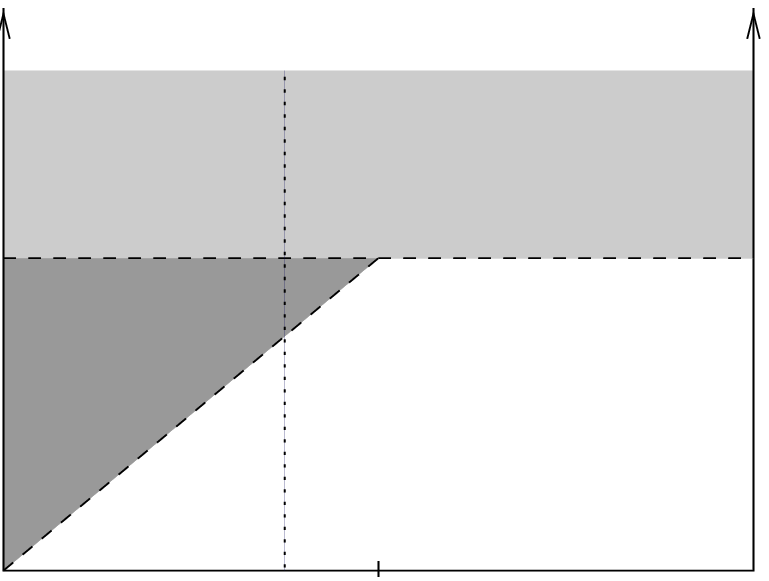, height=6cm}}
\put(300,10){\makebox(0,0){$\phi/\pi$}}
\put(272,0){\makebox(0,0){$1$}}
\put(162,0){\makebox(0,0){$0.5$}}
\put(134,0){\makebox(0,0){$0.4$}}
\put(38,160){\makebox(0,0){$v$}}
\put(38,103){\makebox(0,0){$1$}}
\put(90,80){\makebox(0,0){$B_{\rm local}$}}
\put(100,130){\makebox(0,0){$B_{\rm global}$}}
\put(220,108){\makebox(0,0){$\rho=\infty$}}
\put(200,20){\makebox(0,0){$A/B$}}
\put(141,70){\makebox(0,0){$D$}}
\put(141,110){\makebox(0,0){$C$}}
\put(200,50){\makebox(0,0){$B_{\rm complex}$}}
\end{picture}\caption{Phase diagram of the dominant saddle point.  The saddle point is a
global or local minimum of energy in the two shaded regions. The slice at $\phi=2\pi/5$ is
shown in figure \ref{alls} as an illustration. \label{figpha}}}

We next turn  to the region  $v<1$. Since ${\cal E}^{(B)}(\rho)$ goes
from $-\infty$ to $\infty$, it must now have  an even number of extrema. 
What we  have found numerically are two distinct  regions, 
 separated by a boundary
 which  goes from the point $(\pi/2,1)$ to  the origin in the
$(\phi,v)$ plane
(see figure \ref{figpha}). 
In the left region ${\cal E}^{(B)}(\rho)$ has  two extrema --
 a local minimum and a local maximum, depicted as $B$ and $B'$ in figure \ref{alls} --
which, by continuity, should  govern the behavior of the real and of the imaginary
parts of the amplitude. In this region the absorptive part continues 
to be  exponentially smaller than  the real part. As $v$ approaches
$1$ from below, the local maximum saturates the imaginary part of the
amplitude (point $C$ in figure \ref{alls}), as the corresponding
world-sheet modulus runs away to infinity. The local minimum on the
other hand crosses smoothly the $v=1$ threshold, ensuring a continuous
real part of the amplitude.
At the phase boundary
the two saddle points coalesce (point $D$ in figure \ref{alls}),
before moving  off the imaginary $\tau$-axis. In the right region,
${\cal E}^{(B)}(\rho)$ has thus no real saddle point at $v<1$, and the
amplitude must be governed either by the complex solutions of
\eqref{eq:exp55}, or else by the  configuration A. 

\FIGURE{
\begin{picture}(300,180)(0,0)
\put(0,0){\epsfig{file=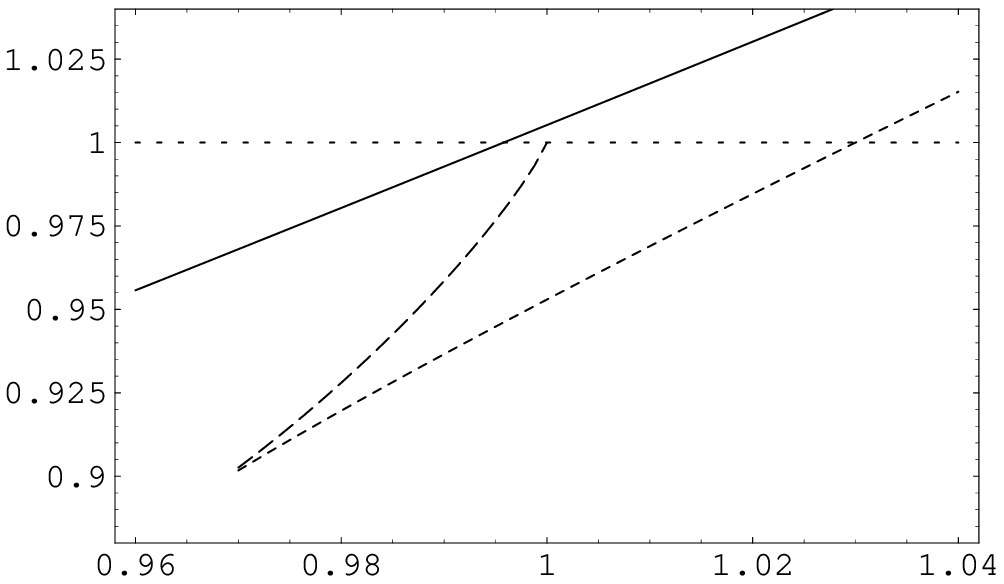, height=6cm}}
\put(295,17){\makebox(0,0){$v$}}
\put(-20,137){\makebox(0,0){${\cal E}/2 s \log 2$}}
\put(50,110){\makebox(0,0){$A$}}
\put(90,80){\makebox(0,0){$B'$}}
\put(120,50){\makebox(0,0){$B$}}
\put(157,121){\makebox(0,0){$C$}}
\put(55,38){\makebox(0,0){$D$}}
\end{picture}\caption{Energy of the deformed saddle points A and B as
a function of the distance $v$ for scattering angle $\phi=2\pi/5$. A
exists at all values of $v$, whereas $B$ starts appearing at
a critical value $v_D(\phi)$ as a pair of local minima and maxima $B$
and $B'$. The latter disappears at $v=1$ ($C$), while the former 
extends smoothly to $v>1$. At $v<v_D$, $B$ and $B'$ move off the
imaginary $\tau$ line. \label{alls}}}

\FIGURE{
\begin{picture}(250,470)(0,0)
\put(0,320){\makebox{\epsfig{file=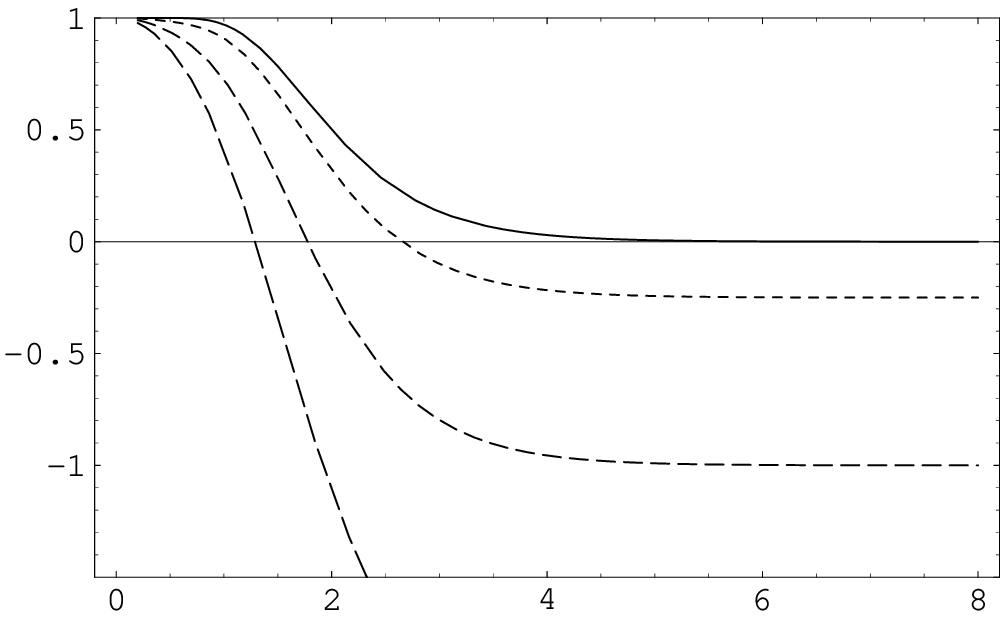,height=5cm}}}
\put(8,160){\makebox{\epsfig{file=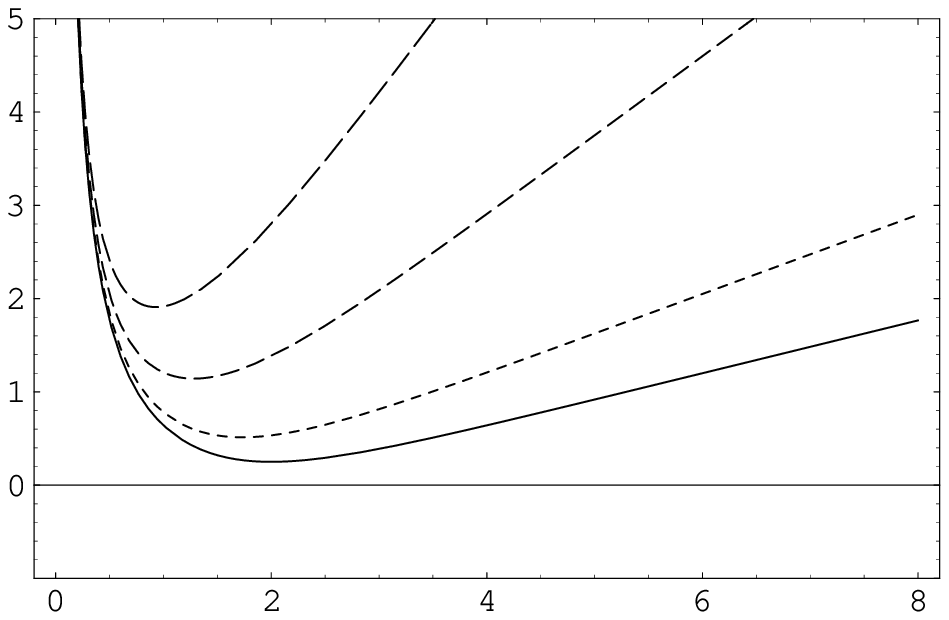,height=5cm}}}
\put(0,0){\makebox{\epsfig{file=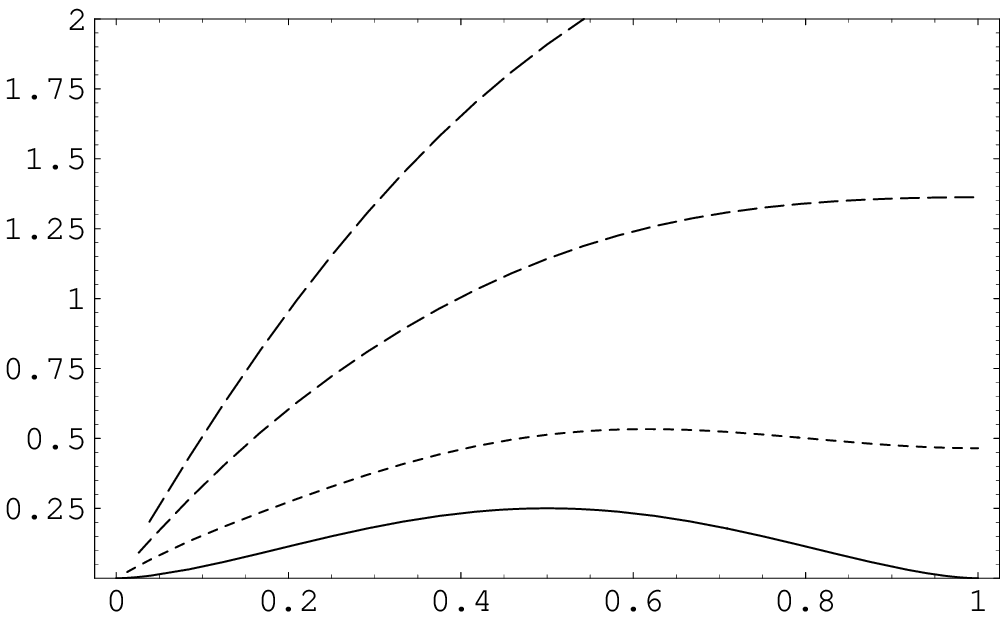,height=5cm}}}
\put(-30,430){\makebox{$\cos^2{\phi\over 2}$}}
\put(-50,276){\makebox{${\cal E}/{2s\log 2}$}}
\put(-60,120){\makebox{$\tilde{\cal E}/{2s\log 2}$}}
\put(240,325){\makebox{$\rho$}}
\put(240,165){\makebox{$\rho$}}
\put(240,5){\makebox{$\phi/\pi$}}
\put(195,417){\makebox{$v=0$}}
\put(195,385){\makebox{$v=.5$}}
\put(195,342){\makebox{$v=1$}}
\put(36,340){\makebox{$v=1.5$}}
\end{picture}
\caption{Deformed saddle point A. Top: $\cos^2{\phi\over 2}$ vs. modular
parameter $\rho$ as in \eqref{eq:exp5}. Middle: electrostatic energy
vs. modular parameter $\rho$ at right-angle scattering. Bottom:
electrostatic energy vs angle. The solid
line corresponds to coincident branes, and the dashed  lines
to separated branes. \label{figGM}}
}

In order to clarify this issue we now turn to an analysis of the
saddle-point configuration  A. In contrast to configuration B which is one of stable
equilibrium on a given surface, this configuration is unstable under
displacements of the positions $z_i$, due to the proximity of
repelling charges. An asymptotic analysis,
on the other hand, shows that  ${\cal E}^{(A)}(\rho)$
has a minimum ${\tilde\rho}$ for
all possible  $(v,\phi)$. We find numerically that this minimum is
unique, and ranges from $0$ to a finite value when $v$ is increased
from zero to infinity, in contrast to the $v=0$ case where $\rho$
runs away to infinity as the angle approaches $\pi$ (see figure
\ref{figGM}). In particular, the solution experiences no discontinuity
at the $v=1$ threshold .

If configuration  A is the dominant saddle-point when  $v=0$, we may 
expect it to remain dominant throughout the lower right region of figure
\ref{figpha}. Our numerical investigation of the solution shows, however, that this
is impossible. The energy ${ \tilde{\cal E}}^{(A)}$ at the phase
boundary is larger than the energy ${ \tilde{\cal
E}}^{(B)}$ (see figure \ref{alls}), implying  an unacceptable exponential jump in the
amplitude. Such an  exponential discontinuity is,
in particular,  incompatible with our
calculation of the absorptive part \eqref{eq:abs}  near
threshold. Thus the dominant saddle point in this region should be the
complex saddle-point of type B. As $v$ goes to zero, the latter
goes over to a modular transform of the saddle point A, consistently with the
hypothesis of \cite{GMa} that A dominates the amplitude when $v=0$.

\clearpage

\setcounter{equation}{0}
\section{Minimal Surfaces and Quantum Tunneling}
The exponential behavior of the amplitudes  is suggestive  (a) of
quantum tunneling, and (b) of a minimal-area problem.  In this section
we will use T-duality to render these two statements  explicit.
Though the mathematics do not change, T-dual language  renders as often
several  aspects of the physics  more transparent.
The key point is that under T-duality 
the fastly-moving  open strings can turn into very long static strings 
stretched between  far-separated D-branes.

Consider in particular a kinematical configuration T-dual 
to right-angle scattering, obtained by dualizing the direction of
outgoing momenta. This amounts to keeping
$ p_1=(E, p ,0),  p_2=(E, -p ,0)$ while 
replacing $p_3$ and $p_4$ by the
T-dual light-like momenta
$ w_3=(-E,0,r/2\pi\alpha^\prime),\ w_4=(-E,0,-r/2\pi\alpha^\prime)$, 
characterizing a pair of static unexcited strings stretching with
opposite orientations between two D-branes. 
The Veneziano amplitude corresponds now to 
 the amplitude for the production of these  static strings in
the collision of two fastly-moving massless quanta. The exponential
behavior  of the amplitude confirms the expectation  that such
processes should be  suppressed by  the small overlap of wavefunctions. 
This interpretation can be used to reproduce the absorptive part of
the one-loop amplitude of section 4.

Another  configuration T-dual 
to right-angle scattering, can be obtained by dualizing all four
external momenta to
$w_1=(E,r/2\pi\alpha^\prime,0),\ w_2=(E,-r/2\pi\alpha^\prime,0),
\ w_3=(-E,0,r/2\pi\alpha^\prime),\ w_4=(-E,0,-r/2\pi\alpha^\prime)$. 
The Veneziano  amplitude is now interpreted as
the amplitude for two oppositely oriented 
such strings to join and flip as depicted in figure
\ref{soap}. This  transition 
requires pulling the two stretched strings so that they join
in the middle, and can then separate  in the perpendicular direction.
The energy barrier due to the string tension, 
$(\sqrt{2}-1)r/\pi\alpha'$, 
is very large in the semiclassical regime under consideration, so
that the  tunneling is exponentially suppressed.
The world-sheet instanton mediating this process corresponds
to the minimal surface interpolating between the world-sheets
of the strings 1 and 2
in the far past and those of 3 and 4 in the far future, 
and can be thought of as a
`soap-bubble' problem for a wire with the contour shown by the thick
line on the left-hand side of figure \ref{soap}. Deforming the square into a
rhombus of angle $\phi$ corresponds to going away from the 
right-angle scattering point.

\EPSFIGURE{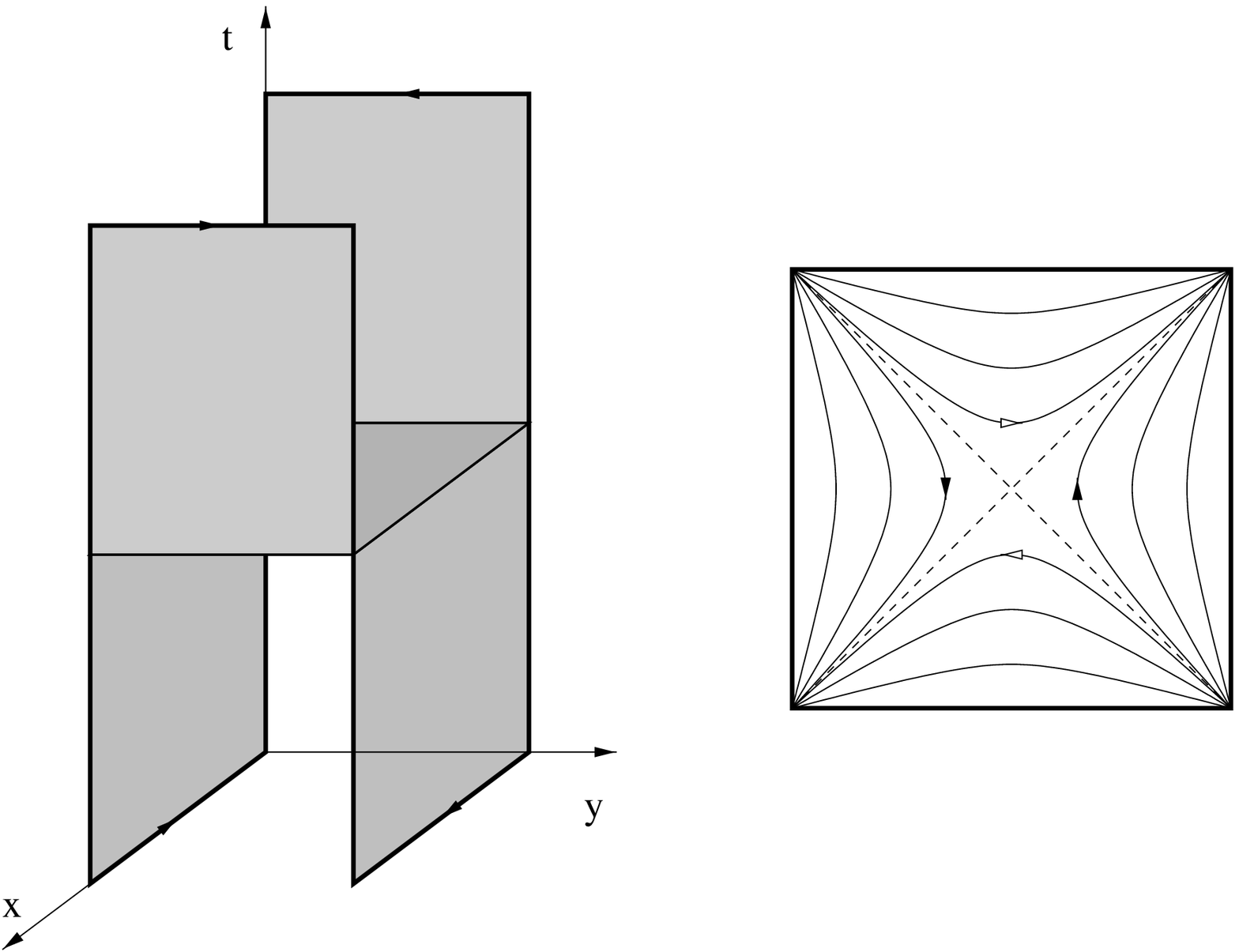,height=8cm}{Disk-level
 high energy scattering in T-dual description.
Left: two very long open strings of opposite orientation
stretched between two pairs of  D-branes undergo a flip transition. 
The heavy line defines  the boundary for the corresponding soap-bubble
problem. Right: time slices of the flip transition. The
dark (resp. light) arrows represent the strings before
(resp. after) the flip.\label{soap}}

The saddle point analysis of high energy scattering actually tells us
what is the precise minimal surface for this soap-bubble problem.
Indeed, for a disk topology, the saddle point
semi-classical configuration is given by \cite{GM,GMa}
\begin{equation}
X^\mu _{\rm saddle}(z) = {i \alpha'} \sum_{i=1}^4 w_i ^\mu  \left(
\log(z-x_i) \pm \log(\bar z - x_i) \right)
\label{sadsol}
\end{equation}
where the $\pm $ sign corresponds to the Neumann or Dirichlet
components respectively. Choosing the insertion points as
$0,1,1/2,\infty$ as implied by \eqref{eq:ven3} for right-angle
scattering, gives
\begin{subequations}
\begin{eqnarray}
X^0(z,\bar z)&=& i {r\over 2\pi}\log {z \bar z (1-z)(1-\bar z) \over 
(z-1/2)(\bar z-1/2)}  \\
X^1(z,\bar z)&=&i {r\over 2\pi} \log {z(1-\bar z)\over \bar z (1-z) } \\
X^2(z,\bar z)&=&i {r\over 2\pi} \log {z-1/2\over \bar z-1/2}  
\end{eqnarray}
\end{subequations}
It is easy to check that this solution satisfies the appropriate 
limits, for instance $X^0\to -i\infty,\ X^1\to r \theta/\pi ,\ X^2\to 0 $
as $z\to 0$, where $\theta\in[0,\pi]$ depends on the direction
in which $z=0$ is approached. The area of the minimal surface is 
given by the saddle point energy $\alpha' s \log 2$ -- this  is  smaller
in particular 
than $\pi \alpha' s/2$ which corresponds  to the  instantaneous transition
of  figure
\ref{soap}. In fact, as emphasized in \cite{GM}, higher genus
world-sheets yield electrostatic energies which are a fraction of the above; 
this implies that the area of the film in the soap-bubble problem
can be reduced by  creating handles. In a realistic setting
this is counterbalanced by  a cost due to the extra curvature  which prevents
the surface from  degenerating  into a foam.

In string theory, the dominance
of higher genus world-sheets implies that perturbation
theory breaks down at high energy, unless the string coupling $g_s$
controlling the curvature cost is tuned to zero
at the same time as the energy is increased \cite{MO}. 
In particular, the contribution of D-instantons should be taken into
account, and may drastically modify 
the exponential softness  of the amplitude.
High-energy scattering in the background of
a D-instanton is known in particular  to be power-suppressed
 \cite{GG},  implying that it is not a tunelling process.
This is  evident in our T-dual picture,
since a D-instanton T-dualizes into an Euclidean D-brane cutting
through figure \ref{soap} at a given time. The open string end-points  can move
freely on this slice, so that there is no energetic barrier to the
flip transition.

\acknowledgments
The authors are grateful to the organizers of the Extended Workshop
on String Theory at ICTP, Trieste, during which this work
was started, for their kind hospitality and support. We also thank
Eric D'Hoker, Michael  Green and Elias  Kiritsis for discussions.
This research is supported in part by the EEC 
under the TMR contract ERBFMRX-CT96-0090.

\vskip .5cm
\noindent {\it Note added in proof.}
After the present work was completed, another work \cite{ah}
appeared which discusses some related issues in the context
of brane-world models, where quark and leptons are 
localized on separated walls. Their treatment is purely
field-theoretical, and gives thus the characteristic   $e ^{-\sqrt{-t} r}$
behavior that we have discussed in section 4. Such a field-theoretical
treatment would be  applicable if the string scale were much heavier than the
(Standard Model) Kaluza-Klein scale, which does not occur in
weakly-coupled type I models
(see however \cite{ap} for such a model in type IIB theory).


\end{document}